\PassOptionsToPackage{unicode}{hyperref}
\PassOptionsToPackage{hyphens}{url}
\PassOptionsToPackage{dvipsnames,svgnames,x11names}{xcolor}
\documentclass[
  12pt]{article}
\usepackage{xcolor}
\usepackage{amsmath,amssymb}
\usepackage{iftex}
\ifPDFTeX
  \usepackage[T1]{fontenc}
  \usepackage[utf8]{inputenc}
  \usepackage{textcomp} 
\else 
  \usepackage{unicode-math} 
  \defaultfontfeatures{Scale=MatchLowercase}
  \defaultfontfeatures[\rmfamily]{Ligatures=TeX,Scale=1}
\fi
\usepackage{lmodern}
\ifPDFTeX\else
\fi
\IfFileExists{upquote.sty}{\usepackage{upquote}}{}
\IfFileExists{microtype.sty}{
  \usepackage[]{microtype}
  \UseMicrotypeSet[protrusion]{basicmath} 
}{}
\makeatletter
\@ifundefined{KOMAClassName}{
  \IfFileExists{parskip.sty}{%
    \usepackage{parskip}
  }{
    \setlength{\parindent}{0pt}
    \setlength{\parskip}{6pt plus 2pt minus 1pt}}
}{
  \KOMAoptions{parskip=half}}
\makeatother
\makeatletter
\ifx\paragraph\undefined\else
  \let\oldparagraph\paragraph
  \renewcommand{\paragraph}{
    \@ifstar
      \xxxParagraphStar
      \xxxParagraphNoStar
  }
  \newcommand{\xxxParagraphStar}[1]{\oldparagraph*{#1}\mbox{}}
  \newcommand{\xxxParagraphNoStar}[1]{\oldparagraph{#1}\mbox{}}
\fi
\ifx\subparagraph\undefined\else
  \let\oldsubparagraph\subparagraph
  \renewcommand{\subparagraph}{
    \@ifstar
      \xxxSubParagraphStar
      \xxxSubParagraphNoStar
  }
  \newcommand{\xxxSubParagraphStar}[1]{\oldsubparagraph*{#1}\mbox{}}
  \newcommand{\xxxSubParagraphNoStar}[1]{\oldsubparagraph{#1}\mbox{}}
\fi
\makeatother

\usepackage{color}
\usepackage{fancyvrb}

\DefineVerbatimEnvironment{Highlighting}{Verbatim}{commandchars=\\\{\}}
\usepackage{framed}
\definecolor{shadecolor}{RGB}{241,243,245}
\newenvironment{Shaded}{\begin{snugshade}}{\end{snugshade}}

\newcommand{\AttributeTok}[1]{\textcolor[rgb]{0.40,0.45,0.13}{#1}}

\newcommand{\ControlFlowTok}[1]{\textcolor[rgb]{0.00,0.23,0.31}{\textbf{#1}}}

\newcommand{\DecValTok}[1]{\textcolor[rgb]{0.68,0.00,0.00}{#1}}

\newcommand{\ErrorTok}[1]{\textcolor[rgb]{0.68,0.00,0.00}{#1}}

\newcommand{\FloatTok}[1]{\textcolor[rgb]{0.68,0.00,0.00}{#1}}
\newcommand{\FunctionTok}[1]{\textcolor[rgb]{0.28,0.35,0.67}{#1}}

\newcommand{\NormalTok}[1]{\textcolor[rgb]{0.00,0.23,0.31}{#1}}

\newcommand{\OtherTok}[1]{\textcolor[rgb]{0.00,0.23,0.31}{#1}}

\newcommand{\SpecialCharTok}[1]{\textcolor[rgb]{0.37,0.37,0.37}{#1}}

\newcommand{\StringTok}[1]{\textcolor[rgb]{0.13,0.47,0.30}{#1}}

\usepackage{longtable,booktabs,array}
\usepackage{calc} 
\usepackage{etoolbox}
\makeatletter
\patchcmd\longtable{\par}{\if@noskipsec\mbox{}\fi\par}{}{}
\makeatother
\IfFileExists{footnotehyper.sty}{\usepackage{footnotehyper}}{\usepackage{footnote}}
\makesavenoteenv{longtable}
\usepackage{graphicx}
\makeatletter
\newsavebox\pandoc@box
\newcommand*\pandocbounded[1]{
  \sbox\pandoc@box{#1}%
  \Gscale@div\@tempa{\textheight}{\dimexpr\ht\pandoc@box+\dp\pandoc@box\relax}%
  \Gscale@div\@tempb{\linewidth}{\wd\pandoc@box}%
  \ifdim\@tempb\p@<\@tempa\p@\let\@tempa\@tempb\fi
  \ifdim\@tempa\p@<\p@\scalebox{\@tempa}{\usebox\pandoc@box}%
  \else\usebox{\pandoc@box}%
  \fi%
}
\def\fps@figure{htbp}
\makeatother

\providecommand{\tightlist}{%
  \setlength{\itemsep}{0pt}\setlength{\parskip}{0pt}}

\usepackage[]{natbib}
\usepackage{booktabs}
\usepackage{longtable}
\usepackage{array}
\usepackage{multirow}
\usepackage{wrapfig}
\usepackage{float}
\usepackage{colortbl}
\usepackage{pdflscape}
\usepackage{tabu}
\usepackage{threeparttable}
\usepackage{threeparttablex}
\usepackage[normalem]{ulem}
\usepackage{makecell}
\usepackage{xcolor}
\usepackage{setspace}
\usepackage{etoolbox}
\AtBeginEnvironment{Shaded}{\singlespacing}
\makeatletter
\@ifpackageloaded{caption}{}{\usepackage{caption}}
\AtBeginDocument{%
\ifdefined\contentsname
  \renewcommand*\contentsname{Table of contents}
\else
  \newcommand\contentsname{Table of contents}
\fi
\ifdefined\listfigurename
  \renewcommand*\listfigurename{List of Figures}
\else
  \newcommand\listfigurename{List of Figures}
\fi
\ifdefined\listtablename
  \renewcommand*\listtablename{List of Tables}
\else
  \newcommand\listtablename{List of Tables}
\fi
\ifdefined\figurename
  \renewcommand*\figurename{Figure}
\else
  \newcommand\figurename{Figure}
\fi
\ifdefined\tablename
  \renewcommand*\tablename{Table}
\else
  \newcommand\tablename{Table}
\fi
}
\@ifpackageloaded{float}{}{\usepackage{float}}
\floatstyle{ruled}
\@ifundefined{c@chapter}{\newfloat{codelisting}{h}{lop}}{\newfloat{codelisting}{h}{lop}[chapter]}
\floatname{codelisting}{Listing}

\makeatother
\makeatletter
\@ifpackageloaded{caption}{}{\usepackage{caption}}
\@ifpackageloaded{subcaption}{}{\usepackage{subcaption}}
\makeatother
\usepackage{bookmark}
\IfFileExists{xurl.sty}{\usepackage{xurl}}{} 
\hypersetup{
  pdftitle={Pivoting the Paradigm: The Role of Spreadsheets in K--12 Data Science},
  pdfauthor={Oren Tirschwell; Nicholas Jon Horton},
  pdfkeywords={Computational Thinking, Data Acumen, Science
Teaching, STEM Education, Data Tools},
  colorlinks=true,
  linkcolor={blue},
  filecolor={Maroon},
  citecolor={Blue},
  urlcolor={Blue},
  pdfcreator={LaTeX via pandoc}}

\begin{document}

\def\spacingset#1{\renewcommand{\baselinestretch}%
{#1}\small\normalsize} \spacingset{1}


\date{June 3, 2026}
\title{\bf Pivoting the Paradigm: The Role of Spreadsheets in K--12 Data
Science}
\author{
Oren Tirschwell\\
Department of Statistics, Amherst College\\
and\\Nicholas Jon Horton\footnote{nhorton@amherst.edu}\\
Department of Statistics, Amherst College\\
}
\maketitle

\bigskip
\bigskip
\begin{abstract}
Spreadsheet tools are widely accessible to and commonly used by K--12
students and teachers. While spreadsheets are not ideal for many types
of statistical analysis, they have an important role in data collection
and organization. From a pedagogical standpoint, spreadsheets make data
visible and easy to interact with, facilitating student engagement in
data exploration, analysis, and computation. Though not suitable for all
tasks, spreadsheets can facilitate learning and practicing data and
computing skills for K--12 students. This paper 1) demonstrates the
potential utility of spreadsheets in K--12; 2) reviews prior frameworks
and standards that are relevant for K--12 data tools; and 3) proposes
data-driven data skills to help develop data acumen and computational
fluency. We provide some example activities, identify challenges and
barriers to adoption, suggest pedagogical approaches to ease the
learning curve for instructors and students, and discuss the need for
professional development to facilitate deeper use of spreadsheets for
data science and STEM disciplines.
\end{abstract}

\noindent%
{\it Keywords:} Computational Thinking, Data Acumen, Science
Teaching, STEM Education, Data Tools
\vfill

\newpage
\spacingset{1.9} 

\section{Introduction}\label{sec-intro}

Data and computing have become instrumental to all facets of life in the
21st century \citep{nasemk12:2026}. While K--12 school districts have,
to varying degrees, begun to adapt their curricula to teach data and
computing related skills to students
\citep{rosenberg_big_2022, nasemwork:2023, dslp, nasemk12:2026}, the
question of how best to foster these skills practically and equitably
remains unsolved.

In this paper, we propose that spreadsheets---despite well-documented
limitations \citep{nashjohn:1996, piment:2022, eric:2022}---can play a
meaningful role in developing K--12 data and computing skills. We argue
that spreadsheets are part of the toolkit students already encounter,
and that their flexibility and prevalence in K--12 settings make them
worth taking seriously as a mechanism for skill development.

Recent reports have emphasized the importance of developing data acumen
and computational fluency for K--12 students \citep{nasemk12:2026}. Data
literacy, which some argue encompasses statistical literacy, can begin
early in schools \citep{gould_2022}. Data acumen is a foundational set
of skills and understandings that an individual should have to
effectively leverage data science \citep{NAP25104}. Computational
thinking facilitates problem solving and algorithmic thinking
\citep{wing_2006, vanb:2023}, and is a critical component of data
science and statistics \citep{nasemk12:2026, pcmi2016guidelines}.

The goal of this paper is threefold:

\begin{enumerate}
\def\labelenumi{\arabic{enumi}.}
\tightlist
\item
  Demonstrate the potential utility of spreadsheets in a K--12 data and
  computing curriculum;
\item
  Describe how widely adopted standards documents already outline
  related data and computing skills that can be taught and reinforced
  using spreadsheets; and
\item
  Suggest a set of proposed skills that complement other data and
  computing learning experiences.
\end{enumerate}

We acknowledge that other data science tools commonly used in K--12
settings (e.g., CODAP, GeoGebra, Desmos, TuvaLabs, and even R or Python)
should serve as complements to spreadsheets, each bringing its own
strengths to data exploration \citep{moon_2023}. Our focus on
spreadsheet tools such as Google Sheets and Excel reflects their
near-universal availability in school districts and the foundational
role they can play in a broader problem solving toolkit (which
complements statistics and data science analysis tools).

Toward these goals, we first evaluate spreadsheets against an existing
framework for data science tools to better assess their advantages and
limitations (Section~\ref{sec-spreadsheets}). We then review relevant
standards and frameworks that inform data and computing skills K--12
students might develop (Section~\ref{sec-standards}). After addressing
\emph{how} K--12 students might develop data and computing skills, we
turn to \emph{what} they should ultimately gain from working with
spreadsheets. In Section~\ref{sec-skills}, we propose five data skills,
each illustrated with sample activities that could be integrated across
a variety of courses (including but not limited to science, math, and
computer science). Finally, we outline challenges, limitations, and the
tangible benefits of teaching with spreadsheets
(Section~\ref{sec-discussion}).

\section{Spreadsheets---A Part of the Data and Computing
Toolbox?}\label{sec-spreadsheets}

Spreadsheets are a common way that data are entered, stored, analyzed,
and visualized \citep{brom:woo:2018}. These tools are widely used in
K--12 settings. A sample of 330 K--12 science educators showed 89\% use
spreadsheet tools with their students \citep{rosenberg_big_2022}. Of
those, the vast majority use Google Sheets---either alone or alongside
Excel, since students and teachers often have licensing by default
\citep{rosenberg_messy, moon_2023}. \citet{Hohenwarter_2014} notes that
K--12 mathematics software tools often combine a spreadsheet component
with a dynamic geometry system and a computer algebra system.
\citet{Hudson03072025} describe how spreadsheets can be used to support
computational thinking and data acumen (e.g., data moves) in secondary
school statistics lessons.

Past work has considered guidelines and best practices surrounding the
use of spreadsheets for data science. For instance,
\citet{hertz_eleven_2024} provide a number of tips to effectively
utilize a spreadsheet tool for analysis, focused on the importance of
making spreadsheets machine-readable (including by having one data point
per cell), using computer-friendly characters, defining null values, and
being thoughtful around datetime records. \citet{brom:woo:2018} share
detailed guidance for how data should be organized in spreadsheets. They
call for consistency, a thoughtful choice of names for columns and
files, a prohibition on calculations within a data spreadsheet, and a
plea that data be stored in a rectangular format \citep{wick:2014}.

Should spreadsheets be used beyond just as a data organization tool? We
believe that they are a component of a broader toolkit and that repeated
learning experiences with spreadsheets will deepen data and computing
skills. To justify this claim we begin by reviewing frameworks for
teaching tools.

There is a considerable history of evaluating the strengths and
weaknesses of teaching tools in statistics, beginning with
\citet{biehler1997}. \citet{mcnamara_key_2018} builds on this prior
structure to describe key attributes of modern statistical computing
tools, relevant for novice and expert users.

\citet{piment:2022} extend and adapt the frameworks proposed by
\citet{biehler1997} and \citet{mcnamara_key_2018} to evaluate K--12 data
science tools. These tools are classified into four categories:
Spreadsheets, Visual Interfaces, Scripting Languages, and Other
Interfaces.

\citet{piment:2022} conclude that ``relying solely on spreadsheets as a
data analysis tool is likely to lead to a frustrating experience for
those learning and engaging in data science.'' While we concur with that
assessment (relating to data analysis), and are not suggesting that
spreadsheets are the sole tool students should utilize, we believe that
spreadsheets have a number of strengths that make them a valuable part
of the data and computing toolbox for K--12 students, particularly in
the areas of data organization, making data visible and interactive, and
fostering computational skills and thinking \citep{wing_2006}.

In this section, we expand upon the strengths and weaknesses of
spreadsheets in K--12 data and computing education. We elaborate upon
each of the elements of the \citet{piment:2022} framework (reprinted in
Table~\ref{tbl-framework}), arguing that while the conclusion of
\citet{piment:2022} has merit in some areas, it understates the
strengths spreadsheets bring in others, particularly computing and data
organization skills, and the ability to make data visible and
interactive for students. For each of the elements, we provide an
informal grade for how well spreadsheet tools address each item in the
framework.

\begingroup
\renewcommand{\baselinestretch}{1}\selectfont

\begingroup\fontsize{9}{11}\selectfont

\begin{longtable}[t]{>{\raggedright\arraybackslash}p{4cm}>{\raggedright\arraybackslash}p{8cm}>{\raggedright\arraybackslash}p{3cm}}

\caption{\label{tbl-framework}Framework for considering K--12 data
science tools, adapted from McNamara's \citeyearpar{mcnamara_key_2018}
key attributes for a modern statistical computing tool. The `Grade'
column represents our informal assessment for how well spreadsheet tools
address each item in the framework. We have split out `Reproducible
workflows' from `Data analysis cycle'. The first two columns are
reproduced with permission from the National Academy of Sciences,
Courtesy of the National Academic Press, Washington, D.C.}

\tabularnewline

\toprule
\textbf{Feature} & \textbf{Description} & \textbf{Grade}\\
\midrule
\endfirsthead
\multicolumn{3}{@{}l}{\textit{(continued)}}\\
\toprule
\textbf{Feature} & \textbf{Description} & \textbf{Grade}\\
\midrule
\endhead

\endfoot
\bottomrule
\endlastfoot
Accessibility & Includes cost, simplicity of cloud-based tools, disability access, multilingual support & A\\
Data as a first-order object & Data as primary interest: hierarchical vs. tabular formats, viewing data; key to building ``students conception of data" & A+\\
Ease of entry & Clarity about how the tool works; includes consideration of students' conceptions of data and developmental appropriateness & A-\\
Data analysis cycle; Reproducible workflows & Iterative cycle of posing questions, exploring data, visualizing results, modeling, model assessment, and communicating results; reproducing data wrangling, analyses, and explorations & C\\
Interactivity & Support for direct interaction with data, e.g., pinch, click-and-drag, brushing, hovering & D / A with other tools\\
Flexible plot creation & Univariate, bivariate, and multivariate displays with ability to augment graphics in a variety of ways & D / A with other tools\\
Inferential analysis & Reasoning with samples and inferring beyond data; support for simulations and resampling; offering probabilistic or uncertain expressions of data & --\\
Non-standard data & Working with multiple forms of data such as spatial data, network data, etc. & C\\
Extensibility & (mostly beyond our scope, but see Google Apps Script) & --\\*

\end{longtable}

\endgroup{}

\endgroup

\subsection{Accessibility}\label{accessibility}

\citet{piment:2022} note that \emph{Accessibility} is a strong suit of
spreadsheets in a K--12 context. We concur. Most school districts
provide access to either Google Sheets or Excel to all students. Sheets
is fully web-based and has associated apps, so it can be used on any
device with an internet connection. These programs can usually be used
with screen readers.

Given the widespread existing adaptation of Google Sheets in many school
districts \citep{rosenberg_big_2022} and its flexibility and power,
especially when used with connectors, we believe this tool is well
poised to assist in the development of data science and computing skills
for K--12 students across the country.

Pre-existing licensing agreements minimize concerns about potential
safety, security, and professional development implications of acquiring
new software \citep{data-security} and leveraging existing
licenses.\footnote{For instance, schools must comply with the Children
  Online Privacy Protection Act and may be looking for companies to sign
  the ``K--12 Education Technology Secure by Design Pledge'' from the
  Cybersecurity and Infrastructure Security Agency.}

\subsection{Data as a first-order
object}\label{data-as-a-first-order-object}

With spreadsheet tools, data are brought to the fore---they are
presented clearly and directly in front of the user, and they are easy
to interact with. This is an example of \emph{Data as a first-order
object} and allows for a rapid development of student conception of
data. \citet{leemojica2022} describe how expert data scientists often
explore data by scrolling through a spreadsheet to understand structure,
outliers, and other features of the data. This is perhaps one of the
strongest features for spreadsheets. This may be the reason why
spreadsheets are so widely used in K--12 settings, as they allow
students to directly interact with data and develop intuition about what
data look like and how they are structured. It may also account for how
other tools (e.g., CODAP, GeoGebra, Desmos) have incorporated
spreadsheet-like data views to allow users to interact with data (see
Section~\ref{sec-othertools}).

\subsection{Ease of entry}\label{ease-of-entry}

By making data visible to users and interactive, spreadsheet tools
foster \emph{Ease of entry}. Getting started with the basics of data
entry is quite easy, with a nontrivial learning curve to get from data
entry to numerical analysis, and an even greater learning curve to be
able to perform more sophisticated statistical analysis. The simplest
functionality can be taught early on, with more advanced skills and
intuition being built over time. This is another relative strength of
spreadsheet tools.

\subsection{Data analysis cycle}\label{data-analysis-cycle}

Taken together, the range of functionality that spreadsheets possess
encompasses components of the \emph{Data analysis cycle}. Work by
\citet{lee_investigating_2022} describes frameworks that outline core
components of the data analysis cycle. They look to a related idea: the
``Data Investigation Process'' (see also Competency 1 of
\citet{nasemk12:2026}). This process begins by framing a problem; then,
an analyst will collect and record data
\citep{witte_2025, witte_data_2025}. Next, these data will be processed,
explored, and visualized \citep{ridgway_2017}. Finally, students can
undertake additional analysis and simple modeling, then communicate
results and findings.

Using a spreadsheet, students in elementary school could examine
individual records, enter data from simple experiments, and develop
basic intuition about what data look like. Middle school students, with
just a few mouse clicks, can create simple charts to visualize their
data. Basic formulas, such as \texttt{SUM()} and \texttt{AVERAGE()}, can
be taught to generate numerical summaries. Later, students can begin to
use increasingly sophisticated functionality that more closely mimics
the Data Moves outlined by \citet{erickson_data_2019} (and extended to
the spreadsheet context by \citet{moon_2023}). Specifically, these Data
Moves include: filtering, grouping, summarizing, calculating,
merging/joining, and making data hierarchical.\footnote{In consideration
  of these major data moves (which are similar to verbs proposed by
  \citet{wick:2014} and discussed in \citet{chance:2015}): the UI-driven
  filter feature can be employed; more complex summary functions can be
  taught for aggregation (including pivot tables to more closely mirror
  the group-by/summarize approach); and transformations and variable
  manipulation can occur in new columns. Additionally, joins can be
  taught using the \texttt{XLOOKUP()} family of functions. We note that
  some important aspects of data wrangling (e.g., hierarchical
  relationships) are challenging to implement using spreadsheet tools
  \citep{eric:2022}.} Finally, some basic modeling is offered by
spreadsheets.

How can students communicate their results? Although they are able to
copy results from a spreadsheet into a written report, this ``cut and
paste'' process is not ideal or recommended \citep{baumer2014}. More
generally, version control \citep{beckman:2021} is an important aspect
of data science and is an area where there is a need for improved tools
\citep{NAP25104}. While recent improvements in Google Sheets version
history allow for a user to manually trace through some granularity of
edits and adjustments, without significant effort and thought, it is
quite challenging to produce a fully \emph{Reproducible workflow}. The
situation is worse for other common spreadsheet tools, which led to our
informal grade of C for this item in the framework.

\subsection{Interactivity and flexible plot
creation}\label{interactivity-and-flexible-plot-creation}

\emph{Interactivity} and \emph{Flexible plot creation} are two places
where spreadsheets themselves fall short of other data analysis tools.
While spreadsheets are commonly used to create graphical representations
of data \citep{prodromou}, they don't have built-in support for
multivariate displays. As another example, there is no built-in way to
make a simple dot plot (a common graph used in early grades to help
students understand distributions, \citet{konold:2007}). (CODAP, on the
other hand, provides interactive and easy-to-use features to produce
these and other visualizations.) \citet{ridgway_2017} provides a review
of tools for visualizing data.

One solution to the problem of generating multivariate displays is
provided by a connector between Google Sheets and Google Data Studio
(formerly Looker Studio). While Data Studio is an additional tool for
flexible and interactive dashboard creation, it is included with any
Google account. It provides support for a wide range of column types,
the creation of custom metrics and fields (``calculated fields''),
multivariate displays, interactive filters, cross-filtering interaction
across charts, and more.\footnote{It is worth noting that Power BI,
  offered by Microsoft, has many similar features to Google Data Studio.
  However, Power BI is only available for free as a desktop download,
  thus making Data Studio more accessible in a school environment.}

Further interactivity and automation, though it adds complexity, can be
developed in the Google Workspace using Google Apps Script, a
JavaScript-based programming language specific to Google. This language
allows tools such as Sheets, Docs, Slides, Calendar, Mail, and more to
be woven together in unique and useful ways. While it may be more
challenging because of the need for more complex coding, engaging in
learning and projects that employ Apps Script could allow students and
teachers to develop more tool sense, understanding the larger ecosystem
of Google and how different software tools can interact.

\subsection{Inferential reasoning and non-standard
data}\label{inferential-reasoning-and-non-standard-data}

Finally, \emph{Inferential reasoning} and support of \emph{Non-standard
data}, while important for data analysis, fall largely outside of the
scope of this paper (and are areas where spreadsheets are less
well-suited). We note, however, that both Google Sheets and Excel have
built-in support for regular expressions (allowing the processing of
text data), and Google Data Studio has robust support for map-based
data. Google Apps Script can also be used to process unstructured data,
such as JSON, XML, or HTML files.

\subsection{Other Related Tools}\label{sec-othertools}

Spreadsheets are by no means the only tool that can be used to develop
data and computing skills for K--12 students. A number of complementary
tools have been proposed and are being used in K--12 settings, including
CODAP \citep{fris:bieh:2021, Hudson03072025}, GeoGebra
\citep{geogebra, Mousoulides2011, prodromou}, Desmos
\citep{bourassa:2019}, TuvaLabs, and even R or Python \citep{moon_2023}.
Some of these tools feature spreadsheet-like interfaces and data views.
\citet{fris:podw:2022} describe how CODAP and TinkerPlots use
spreadsheet-like tables in effective ways. \citet{fris:bieh:2021} raise
important questions about choice of tools (e.g., educational software
such as CODAP, spreadsheets, or professional programming tools). While
we focus our attention on how data and computing skills can be
facilitated using spreadsheets, we believe that such tools are only a
part of a broader set of tools and technologies that students in K--12
should encounter, and that educators should not adopt a
``one-size-fits-all'' approach to data and computing solely using
spreadsheets.

\section{Data and Computing Standards and
Frameworks}\label{sec-standards}

Having covered \emph{why} we believe spreadsheets (and Google Sheets in
particular) have a role in data and computing, we now describe
\emph{what} we believe these students might learn. We begin by reviewing
relevant standards and frameworks that relate to the teaching of
statistics, mathematics, data science, and computing as a way of
demonstrating that related topics are already part of the K--12
curriculum, albeit with varied uptake.

\subsection{``K--12 Computer Science
Standards''}\label{k12-computer-science-standards}

In 2017, the Computer Science Teachers Association (CSTA) updated their
\emph{K--12 Computer Science Standards}, a ``core set of learning
objectives designed to provide the foundation for a complete computer
science curriculum and its implementation at the K--12 level.''
\citep{csta2017} Their framework lays out high-level concepts (such as
``Computing Systems,'' ``Networks \& The Internet,'' or ``Data \&
Analysis''), as well as sub-concepts (e.g., within ``Data \& Analysis,''
one finds ``Storage,'' ``Collection, Visualization, \& Transformation,''
and ``Inference \& Models''). This structure is meant to develop an
enduring understanding of principles and best practices for working with
computers. Some ideas from the concepts ``Data \& Analysis'' and
``Algorithms \& Programming'' are relevant to developing data acumen and
computing skills and can be fostered by learning to use a spreadsheet.

``Data \& Analysis'' includes ``Collection, Visualization, \&
Transformation'' of data, ranging from the basics of data collection and
simple visuals in elementary school to data transformation and
processing in middle and high school \citep{witte_2025}. The ``Inference
and Models'' subconcept (while less relevant to a spreadsheet-driven
activity) also speaks to the importance of using data and visualizations
to find patterns, communicate them, and predict potential future
outcomes.

The ``Algorithms \& Programming'' subconcept refers to variables as
being ``like a container with a name, in which the contents may change,
but the name (identifier) does not.'' This idea could be understood as
quite similar to a spreadsheet cell: when you use a cell reference
(e.g., \texttt{A2}), you can change the content of that cell, and the
reference will return whatever value is currently in it.

Finally, the idea of ``Modularity'' is useful for the data analysis
process, by decomposing an analysis into a ``precise sequence of
instructions.'' In the real world, one is often presented with data and
asked a question; it is then the job of the analyst to figure out how to
identify (and document) steps to utilize the data to answer that
question.

\subsection{Data Science Learning
Progressions}\label{data-science-learning-progressions}

The Data Science Learning Progressions \citep{dslp} is a recent
framework for K--12 data science education that includes a set of
learning progressions across five strands: ``Dispositions and
Responsibility'', ``Creation and Curation'', ``Analysis and Modeling
Techniques'', ``Interpreting Problems and Results'', and ``Visualization
and Communication''.

Several of the data skills outlined in Strand C (Analysis and Modeling)
reference spreadsheet tools, including grades 3-5 C.2.3b ``use
spreadsheets to visualize trends and relationships'', C.2.3c 3-5.C.2.3c
``use no-code or low-code data science tools. e.g., CODAP, Desmos,
Google sheets'' and grades 6-8 C.5.2b ``use simple mathematical or
computational models (e.g., statistical summaries, spreadsheet formulas)
to describe patterns and relationships in data''. In Strand D
(Interpreting Problems and Results) in grades 9-10 D.1.3b students
``solve a real-world comparison problem using a digital spreadsheet,
such as selecting insurance policies or entering different lotteries.''

\subsection{Standards for K--12
Mathematics}\label{standards-for-k12-mathematics}

\begin{figure}

\centering{

\pandocbounded{\includegraphics[keepaspectratio]{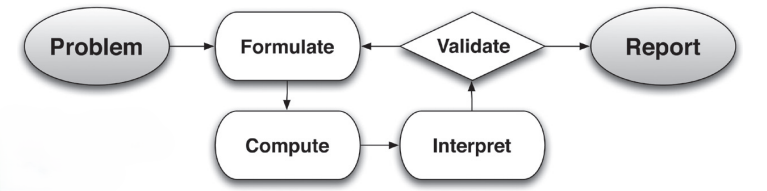}}

}

\caption{\label{fig-modeling-diag}Workflow of modeling from the Common
Core Standards for Mathematics.}

\end{figure}

According to the Common Core Standards for Mathematics
\citep{common-core-2023}, ``Modeling is the process of choosing and
using appropriate mathematics and statistics to analyze empirical
situations, to understand them better, and to improve decisions.'' They
provide Figure~\ref{fig-modeling-diag} to demonstrate the standard
workflow used to create a model. The National Council of Teachers of
Mathematics ``High School Mathematics Reimagined, Revitalized, and
Relevant'' report \citep{nctm2024} reiterates the importance of
mathematical modeling.

Models can be as simple as writing a total purchase price as a product
of the cost per item and the quantity purchased. They can be more
complex, as well: modeling financial data, selecting driving routes,
predicting the weather, and more. Several potential applications of
modeling exist that can be demonstrated using a spreadsheet.

To demonstrate an important concept in mathematics of function
composition (applying a function to the result of a second function,
i.e., \((f \circ g)(x) = f(g(x))\)), students can use formulas to
represent different functions across multiple columns. Properties such
as function associativity and lack of commutativity can be explored and
visualized with real numbers.

Sequences and series provide another straightforward example of how one
might integrate mathematical modeling and spreadsheets. Formulas for
recursive and explicit expressions of geometric series could be
introduced. Consider a scenario where someone deposits \$1,000 into a
bank account each year and accrues annual interest at a rate of 5\%.
They can calculate the total balance of the account at the end of 50
years \citep{geom_series_applications}. This task could first be
explored recursively using a spreadsheet; this may then help students
visualize the series component. Students can then derive an explicit
formula and confirm it matches. More guidance about mathematical
modeling in K--12 can be found in the Guidelines for Assessment and
Instruction in Mathematical Modeling Education (GAIMME) report
\citep{gaimme}.

\subsection{Guidelines for Assessment and Instruction in Statistics
Education}\label{guidelines-for-assessment-and-instruction-in-statistics-education}

The GAISE (Guidelines for Assessment and Instruction in Statistics
Education) II K--12 report outlines skills that students need to make
sense of data \citep{gaise_ii}. The report highlights the importance of
the statistical problem-solving process, the process of collecting and
wrangling data, and the important role of technology in data analysis.

As part of the GAISE II K--12 framework, Level B ``Collect Data/Consider
Data'' includes understanding ``that data are information collected and
recorded with a purpose and can be organized and stored in a variety of
structures (e.g., spreadsheets)'' \citep[pp.~17]{gaise_ii}. The report
also discusses how students need to ``interrogate'' the data to learn
about data production, types of variables, and how data are represented.

\subsection{A Framework for K--12 Science
Education}\label{a-framework-for-k12-science-education}

The \citet{national_2012} published a framework for science education
for K--12 students that relates closely to the data analysis cycle
described in \citet{gaise_ii} and formalized by
\citet{lee_investigating_2022} for the ``Data Investigation Process''.

Chapter 3 notes the importance of ``developing students' knowledge of
how science and engineering achieve their ends while also strengthening
their competency with related practices.'' They focus on practices they
see as both relevant and historically underemphasized in typical science
education: modeling, writing explanations, and ``engaging in critique
and evaluation.''

The third practice listed in Chapter 3 is ``Planning and Carrying Out
Investigations.'' It is suggested that from an early age, students
should be led to think about how data are collected and recorded in
scientific investigations.

The fourth practice listed in this chapter is, ``Analyzing and
Interpreting Data.'' Students need to be able to use spreadsheets,
databases, and other data tools to store and analyze data and
relationships between variables.

Finally, the eighth practice is ``Obtaining, Evaluating, and
Communicating Information.'' Students are encouraged to ``write accounts
of their work, using journals to record observations, thoughts, ideas,
and models. They should be encouraged to create diagrams and to
represent data and observations with plots and tables, as well as with
written text, in these journals.'' This goal is consistent with the use
of a lab notebook and the by-hand recording of data, information, and
reflections. Ideally, technology tools could replace these written
recordings to facilitate improved reproducibility
\citep{baumer2014, nasem2019}.

\subsection{Developing Competencies for the Future of Data and
Computing: The Role of
K--12}\label{developing-competencies-for-the-future-of-data-and-computing-the-role-of-k12}

A recent report from the National Academies provides a set of data and
computing competencies for K--12 data and computing education
\citep{nasemk12:2026}. The report proposes a set of seven competencies
that all K--12 students should develop, beginning in elementary grades:

\begin{enumerate}
\def\labelenumi{\arabic{enumi}.}
\tightlist
\item
  Problem Posing and Problem-Solving Processes
\item
  Producing and Working with Data
\item
  Abstraction, Algorithmic Thinking, and Automation
\item
  Probabilistic and Inferential Reasoning
\item
  Models and Representations
\item
  Technology and Society
\item
  Data and Computing Systems.
\end{enumerate}

A number of these competencies are relevant to the use of spreadsheets
in K--12 education, including ``Producing and Working with Data''
(Competency 2), ``Abstraction, Algorithmic Thinking, and Automation''
(Competency 3), and ``Data and Computing Systems'' (Competency 7). The
report highlights ways that spreadsheets can be a useful way for K--12
students to structure and organize data, particularly early in their
interactions with data.

\subsection{Park City Report and Data Science for Undergraduates:
Opportunities and
Options}\label{park-city-report-and-data-science-for-undergraduates-opportunities-and-options}

At the undergraduate level, both the Park City report
\citep{pcmi2016guidelines} and the National Academies ``Data Science for
Undergraduates'' report \citep{NAP25104} provide guidance for data
science programs. The definition of ``data acumen'' in the ``Data
Science for Undergraduates'' \citep{NAP25104} consensus study report
includes numerous concrete goals that may be relevant at a K--12 level.
Chapter 2 of the report explores the foundational set of skills and
understandings that someone should have to effectively leverage data
science. Data acumen comprises a number of concepts, including
``statistical foundations'' (e.g., exploratory data analysis), ``data
management and curation'' (e.g., data preparation), ``data description
and visualization'' (e.g., visualizations). The section on
``computational foundations'' places emphasis on student ability to ``be
able to learn new data technologies.''

To that end, we believe that developing the ability to self-source
solutions to problems that arise and understanding the reasons for data
collection, storage, and manipulation/wrangling is more important than
just following steps to use a specific spreadsheet tool.

\subsection{Dana Center Course
Frameworks}\label{dana-center-course-frameworks}

Parallel to this research base, the Dana Center's course frameworks
emphasize that all students should be able to critically interpret and
engage with data, attend to data ethics, and use technology to conduct
investigations in authentic contexts \citep{danastats}. The Data Science
framework in particular includes as a learning outcome, ``Access
relevant online datasets/data information (i.e., via spreadsheet imports
or API calls).'' There is also emphasis on the importance of using
``technology to wrangle, visualize, and explore data to develop
conceptual understanding.''

\section{Data and Computing Skills}\label{sec-skills}

Are there ways to support students' movement across an investigation
process---from data creation/collection and structuring, to
exploration/visualization and communication---rather than treating tool
use as a set of isolated procedures?

The standards and frameworks introduced in Section~\ref{sec-standards}
provide guidance for the data and computing skills K--12 students might
find valuable and indicate where in the existing and future curriculum
these skills might be developed. We believe that spreadsheets are a
vehicle through which one can teach general concepts that will build
data and computing skills that complement other data tools benefiting
students into the future.

For instance, a student should understand \emph{why} they would choose a
specific spreadsheet setup to record data from a given experiment, or
\emph{why} they might represent a relationship with a scatterplot rather
than a line graph. Such experiences may be more valuable than a
``cook-book'' approach where they exclusively follow detailed
instructions on \emph{how} to create these outputs given a certain type
of input.

In this section, we propose a set of five aspirational data and
computing skills derived from existing learning standards and frameworks
for undergraduate and K--12 students. We describe how these skills might
be introduced or practiced with a set of sample activities and how they
connect with existing data and computing standards and frameworks.

\begin{enumerate}
\def\labelenumi{\arabic{enumi}.}
\tightlist
\item
  Data entry, aggregation, and visualization
\item
  Data consistency checking
\item
  Mathematical applications within spreadsheets
\item
  Algorithmic thinking and implementation
\item
  Scripting, automation, and tool integration
\end{enumerate}

We now examine each of these data skills in turn and propose several
potential ways in which spreadsheets could be integrated into the K--12
curriculum.

Given the connected applications offered by Google and the added benefit
of Google being entirely web-based (thus standardizing the user
experience across different computer types), Google Sheets is a common
spreadsheet software environment of choice for K--12 districts. For the
remainder of this paper, spreadsheet functions will be explained or
demonstrated as they are written in Google Sheets (small modifications
are required to implement these examples in Excel).

\subsection{Data entry, aggregation, and
visualization}\label{sec-skill1}

The merits of a spreadsheet using data as a first-order object become
immediately clear when simple data entry is employed. This could happen
in a middle school class where students are conducting a science
experiment, estimating the value of \(\pi\) in pre-algebra, or learning
the basics of probability with a coin flip experiment. Instructors can
demonstrate this process on a projector, and students can follow along
themselves.

These types of activities easily lend themselves to demonstrating
arithmetic operations, as well as basic aggregation functions, such as
\texttt{SUM()} or \texttt{AVERAGE()}. In a middle or high school science
lab, once data have been collected, graphical displays can be created to
visualize results \citep{gaise_ii}. This could be anything from scatter
plots to understand force in physics to bar charts comparing average
plant growth under different light conditions in a high school biology
class \citep{national_2012}.

Further, science classes could have individual or small groups of
students collect their own experimental results and then pool these
results across a full class \citep{roadless}. Those who teach the same
class for multiple years could even pool results across time. This would
allow students to better understand sampling error and develop intuition
on how the Law of Large Numbers (the distribution of means is less
variable than the distribution of individual observations) works. Such
spreadsheet activities could complement work done using applets or
similar tools \citep{applets2012, gaise_ii}.

Science and math classes are not the only ones that can develop this
skill. In one 9th-grade English class, students who read \emph{The
Catcher in the Rye} worked in small groups to identify textual evidence
of Holden Caulfield's depressive symptoms, recording results in a
spreadsheet. The teacher aggregated these results and made a bar graph
to spark a data-driven discussion on the severity and context of these
behaviors \citep{ssht-classroom-ex}.

We now present two concrete examples of demo spreadsheets, created to
illustrate how one might go about exploring this first data skill. Each
is an actual Google Sheet that can be accessed via the supplemental
materials. (Both rely on fabricated names and data to avoid privacy
concerns.)

\subsubsection{Paper Airplanes}\label{paper-airplanes}

\begin{figure}

\centering{

\pandocbounded{\includegraphics[keepaspectratio]{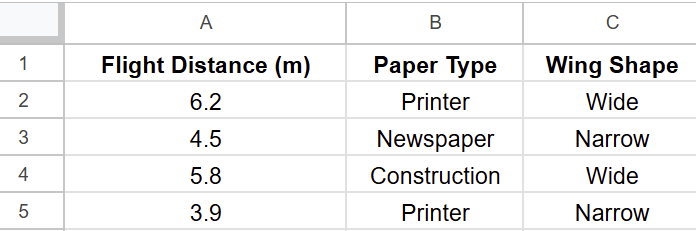}}

}

\caption{\label{fig-flight-demo}Snapshot of header row and first four
rows of data in the ``Data'' tab for the first example spreadsheet.}

\end{figure}

This first example provides data one might collect from an experiment on
paper airplanes, measuring the distance a plane flies considering two
different wing structures and three different types of paper
\citep{boger}. The data itself could be experimentally collected by
students, and data entry into a spreadsheet of this form can be
utilized. A snapshot of the data sheet is presented in
Figure~\ref{fig-flight-demo}. We believe that this form of spreadsheet
use is quite common and facilitates teaching several of the
\citet{nasemk12:2026} competencies (including ``Problem Posing and
Problem-Solving Processes'' and ``Producing and Working with Data'').

A number of summary visualizations and numerical summaries may be
included as part of such an investigation. For instance, perhaps a
student could create a histogram of flight times. Or, summary statistics
aggregated by paper type could be derived. The generation of metrics by
paper type (e.g., average flight distance by paper type) can be
efficiently implemented using a pivot table. A bar chart could be used
to display comparative results (see the ``Results'' tab).

It would be straightforward to move from a spreadsheet tool used for
data collection and organization to a more flexible and interactive
visualization tool, such as CODAP. Such an approach would allow students
to compare and contrast how effectively tools are able to solve
particular tasks \citep[see Competency 7 ``Data and Computing Systems''
from][]{nasemk12:2026}.

\subsubsection{Library}\label{library}

The second example, representing a very small school library, provides
some more sophisticated conceptual functionality. The first tab is a
database of books in the library while the second contains one row for
each time a student checked out one of these books. Book IDs are used to
uniquely identify each book. Thus, consistent with a relational
database, rather than using a book name associated with each checkout
record, we incorporate a book ID. These data are then used, in
conjunction with a \texttt{VLOOKUP()} formula, to automatically populate
the book title in column D.

We provide a number of analytic questions a user might seek to answer
using these data, incorporating different types of aggregation and
exploratory numerical analysis. These questions, quantitative answers,
and the formula used to obtain these answers are available on the tab
titled ``Questions.'' For instance, a simple \texttt{AVERAGE()} formula
is used to compute the mean number of pages per book in the database. An
\texttt{AVERAGEIF()} formula is used to compute the average length of
Fantasy books in the database; this average is higher than the overall
average number of pages per book. Finally, the tab titled ``Checkouts
Pivot Table'' demonstrates an application of a pivot table, specifically
for tracking the number of times each book has been checked out.

The library activity reinforces competencies ``Producing and Working
with Data'' and ``Abstraction, Algorithmic Thinking, and Automation''
from \citet{nasemk12:2026}, as well as the ``Data \& Analysis'' and
``Algorithms \& Programming'' concepts from the CSTA K--12 Computer
Science Standards \citep{csta2017}. It also leverages some of the
hierarchy concepts central to \citet{eric:2022}.

\subsection{Data consistency checking}\label{sec-skill2}

Real-world data analysis requires that someone looking at raw data
possess tools to recognize where that data may be inaccurate or impact
analysis, and take steps to remedy that. This component of data acumen
is often overlooked in K--12 education, but it is an important one that
is included in \citet{nasemk12:2026} as part of the ``Producing and
Working with Data'' competency. For instance, suppose a student
downloads a data source with a numerical column, but the numbers have
commas in them. Subsequent use of functions (e.g., \texttt{SUM()} for
aggregation) may not work properly.

The basics of data cleaning and consistency checking might be introduced
in several ways. First, a teacher could use an artificial example of
fabricated data to demonstrate some typical real-world problems.

Alternatively, real-world data can be used. As an example, a pre-algebra
class or algebra class could model real-world data, allowing the
students to download potentially messy data and work towards cleaning
it.

As a larger scale example, students could take the Census at School
survey \citep{census-school} and then download data from students all
across the country who have also taken the survey. This latter approach
provides a personal connection to the data being analyzed: the students
will themselves know what the questions are and which questions are most
interesting to them. Such approaches are valuable for students to
develop a mindset of data wrangling and to begin to inculcate defensive
coding techniques \citep{McNamara02012018}.

\subsection{Mathematical applications within
spreadsheets}\label{sec-skill3}

High school math classes such as Algebra II and Pre-Calculus contain
numerous mathematical concepts that can be implemented using
spreadsheets. For instance, a unit on exponential growth might cover
compound interest formulas, which students could compare to simple
interest. Such basic modeling is at the core of \citet{nasemk12:2026}'s
``Models and Representations'' competency.

A spreadsheet to demonstrate this model could be made fully interactive,
allowing students to understand some more sophisticated spreadsheet
features: perhaps principal amounts and interest rates can be modified,
frequency of interest payments can be adjusted, etc. This type of
``what-if'' analysis is a powerful way to develop intuition about how
different parameters impact the growth of an investment. Such
applications are straightforward to implement in a spreadsheet and were
some of the earliest use cases \citep{visicalc}.

Larger scale projects that may be appropriate for these classes could
include further financial and mathematical modeling concepts, such as a
project comparing the cost of purchasing and maintaining a gas car
compared to an electric car. The cost of home ownership could be
compared using a renting and a purchasing structure. In both of these
examples, graphs can be used to display findings, with students
encouraged to make these displays as interactive as possible. For
example, a user might be able to adjust the initial parameters and, if
formulas are properly implemented, see those changes permeate through
the spreadsheet.

Finally, units on the basics of probability, at the center of
\citet{nasemk12:2026} competency on ``Probabilistic and Inferential
Reasoning'' present an attractive opportunity to incorporate
spreadsheets into the classroom. Expected value calculations and
randomness are two extremely useful applications. The recent film
\emph{Jerry and Marge Go Large} depicts a couple who found a lottery
with a positive expected value under certain conditions
\citep{frankel_jerry_2022}. Students can use spreadsheets to show the
positive expected value and then be provided with a project or activity
to assess this on a lottery of their own choosing. Later explorations
could use applets or other technologies. Such simulations may help
students explore how expected values can be misleading when not paired
with measures of variability \citep{buffett}.

\subsection{Algorithmic thinking and implementation}\label{sec-skill4}

Spreadsheets provide an opportunity to deepen computational thinking
skills \citep{vanb:2023}. After multiple exposures and opportunities to
practice, students might find themselves implementing more complex,
algorithm-based formulas in Google Sheets. For instance, suppose a
student makes a spreadsheet to store their grades for the class. In cell
\texttt{A1}, the student records a numerical grade; in cell \texttt{A2},
they seek to implement a formula for the letter grade associated with
their numerical grade. Such a formula is fairly convoluted if created
using basic Google Sheets functions---it requires multiple \texttt{IF}
formulas nested within one another. For instance, such a formula for
only letter grades A, B, C, and fail might read:
\texttt{IF(A1\ \textgreater{}\ 90\%,\ "A",\ IF(A1\ \textgreater{}\ 80\%,\ "B",\ IF(A1\ \textgreater{}\ 70\%,\ "C",\ "Fail")))}.
Implementing this type of nested logical structure in code reinforces
\citet{nasemk12:2026} competency ``Abstraction, Algorithmic Thinking,
and Automation'' and related concepts \citep{vanb:2023}.

In 2009, the programming language Google Apps Script was released
\citep{noauthor_launched_2009}. Apps Script is a lightweight,
JavaScript-based scripting language allowing for the automation and
connection of multiple tools within Google Workspace (e.g., Drive,
Sheets, Docs, Calendar, Gmail, etc.). The scripting language is
well-documented, and being JavaScript-based, would be accessible to high
school students with prior programming experience
\citep{bastug_effect_2017}. Additionally, Apps Script is freely
available with any Google account.

While not all K--12 students may need to learn to program or code, and
while there is a need for professional development to support its use,
Apps Script provides an accessible jumping off point for those who
choose to do so.

One powerful application of Google Apps Script, suitable for a basic
introduction to the purpose of the language, is coding custom functions
for a Google Sheet \citep{apps_script_quickstart}. We return to the
prior example of a grading spreadsheet. The Apps Script implementation
of this formula might include \texttt{if} and \texttt{else-if}
statements or a \texttt{switch} statement. Example source code is
provided in \hyperref[appendix-a]{Appendix A} (Example 1).

With examples like these (and with increasingly sophisticated use cases
over time), students can be introduced to more sophisticated algorithmic
thinking and its applications to data.

\subsection{Scripting, automation, and tool
integration}\label{sec-skill5}

As students dive deeper into Google Workspace, they may begin to
encounter problems that can't be solved using the built-in
functionality. This provides an opportunity to deepen their
understanding of the ecosystem of tools and how they can be integrated,
as well as to develop skills in scripting and automation
(\citet{nasemk12:2026} Competencies 3 ``Abstraction, Algorithmic
Thinking, and Automation'' and 7 ``Data and Computing Systems''.)

For instance, suppose a student creates a note-taking template for a
certain class within Google Docs, and they want to make one copy for
each day of the week, renaming them accordingly. Google Drive doesn't
even allow for bulk copying of documents, let alone dynamic re-naming.
Example source code to run this simple automation is provided in
\hyperref[appendix-a]{Appendix A} (Example 2).

As a more powerful and complicated example, perhaps a student creates a
sign-up form for an event RSVP they are co-hosting with another student.
However, this event only has space for 20 attendees. Thus, the two
students wish to receive a notification once they reach 10, 15, and 20
sign-ups, so they can track their progress towards capacity but not have
to manually check a response spreadsheet. Associated code is provided in
\hyperref[appendix-a]{Appendix A} (Example 3).

As a third example, let's suppose a student sets up a spreadsheet as a
job application tracker. This may have a column for the company they are
applying to, a second column for the application deadline, and a check
box for whether they have submitted the application. One week before the
application is due, it might be useful to get an email notification
stating that their deadline is coming up, but only if they haven't
submitted the application already. As before, code to perform such
actions is provided in \hyperref[appendix-a]{Appendix A} (Example 4).

These computing skills, consistent with those that are now commonly
taught in many districts and states \citep{ds4e_state}, could be
developed at a basic level for all students, with potential extensions
and deeper exploration for students with more specialized interests.

All three of these examples reflect problems that can't be solved by
default within Google tools, but Apps Script can and does allow for them
to be accomplished, leveraging the earlier investment in use of Google
Sheets. Activities along these lines could provide students with
practical examples of the uses of Apps Script, tangible use cases for
programming, and a powerful, eye-opening investigation into the
potential for integration of multiple tools.

The integrations that can extend the power of Google Workspace go beyond
just programming. Google Data Studio is a free, web-based drag-and-drop
data visualization software that allows any user with a Google account
to create interactive, permissioned dashboards. It includes a built-in
connector to Google Sheets, and also allows for the easy upload of Excel
files and CSVs. Thus, although Google Sheets itself may lack the
qualities of interactivity and flexible plot creation (mentioned above),
by combining Sheets with Data Studio, both of these features are readily
accessible and can be implemented at a high level very quickly. Such
tools might be used to complement more traditional K--12 data analysis
tools such as CODAP and provide structure to deepen computational
thinking skills \citep{csa2z}.

\section{Discussion and Conclusion}\label{sec-discussion}

Nearly 20 years ago, \citet{konold2007} asked: what are the core ideas
in statistics in data analysis, what statistical capabilities do
citizens need, and ``how do we start early in young people's lives to
develop these capabilities.'' The \citet{nasemk12:2026} K--12 data and
computing competencies report called for a comprehensive rethink of how
K--12 students' learning about data and computing could be integrated
more effectively and begun much earlier.

We believe that spreadsheets have a role in K--12 data and computing
education and can serve as a foundation for work in other tools. They
are not the only tool that should be used to develop data and computing
skills, given the many weaknesses that they have for analysis,
visualization, reproducibility, and modeling (to name just a few). But
they have a number of features that make them particularly well-suited
for this purpose (e.g., bringing data to the fore, ease of use,
interactivity, etc.), are already widely used in K--12 settings, and are
an important workplace tool.

Our aim in this paper is deliberately focused. We do not set out to
build a comprehensive K--12 curriculum or provide empirically validated
lesson plans; rather, we hope to provide some examples that could allow
others to lay conceptual groundwork on which such efforts can build.

We believe that students can achieve deeper data and computing skills
through repeated exposure and thoughtful integration of spreadsheets
across a K--12 curriculum. While spreadsheet use could theoretically be
woven into almost any class \citep{nasemwork:2023, nasemk12:2026}, it
may be most straightforward to integrate them across science, math, and
computer science courses.

Given a fast-changing technology landscape, foundational concepts may be
more useful than knowledge of any specific technology. Learning both the
``how'' and the ``why'' allows spreadsheets to be a vehicle through
which one can teach general concepts, building skills to benefit
students into the future \citep{NAP25104}. We believe that the approach
we suggest here has potential to help integrate and leverage data and
computing into the K--12 curriculum.

In their framework for statistics courses, \citet{danastats} suggest
that students ``use technology and statistical software and applications
that are freely available, easy to use, accessible across platforms and
devices, and robust to multiple contexts and data set sizes, with
ongoing accessibility to students after the class ends.'' Tools such as
Google Sheets meet these criteria in many districts due to existing
licensing.

While we have described how spreadsheets could help develop data and
computing literacy, fluency, and acumen in K--12 students, they are
still subject to a number of notable limitations. A key challenge for
spreadsheets is how they fall short on portions of the data analysis
cycle. For instance, Google Sheets does not have the default ability to
create multivariate scatter plots (though Google Data Studio provides
this feature, albeit at the cost of introducing more complexity and an
additional tool). Additionally, while spreadsheets may be part of a
reproducible workflow, they don't possess strong default features to
allow for the reproducibility of an analytic workflow. Tools better
equipped for a reproducible workflow (e.g., quantitative analysis
programming languages like R) do exist but feature a steep learning
curve.

While spreadsheets can be beneficial in transitioning students from
working with small data sets to medium-sized data sets, they have some
limitations when handling ``big data'' as the term is typically used in
the contemporary context. Specifically, Google Sheets has a maximum of
10 million cells allowed per spreadsheet. More importantly, Google Sheet
functionality and performance degrade as the number of cells and
formulas in a spreadsheet increases \citep{rahman_benchmarking_2020}.

Numerous conversations between the authors and selected K--12 educators
shed further light on pedagogical considerations surrounding the use of
spreadsheet tools within these curricula. One of the biggest challenges
may be a lack of teacher knowledge: teachers who don't themselves know
how to use spreadsheets will likely be reluctant to teach them in
classes, and/or may get frustrated with trying to teach these
technologies to students and give up on them \citep{rosenberg_messy}.
Thus, robust professional development is needed to prepare teachers for
this type of work with students \citep{nasemk12:2026}. These issues are
important areas for future work.

In addition to a lack of teacher knowledge, practitioners often reported
how little students themselves know about spreadsheet tools---even in
high school. Starting from the very basics is crucial. One instructor
has turned formula writing into a verbal call-and-response process for
students. He prompts students: ``What do we start with?'' They are
expected to respond by saying, ``Equals sign.'' Then, a prompt of ``How
do we save our formula?'' is replied to with, ``Click return.''

It is difficult, though important, to help students go from a ``cookie
cutter'' plug-and-chug approach to solving problems more dynamically.
For instance, students may struggle to figure out how to jump over
hurdles they face in the process of working with spreadsheets. There may
be ways for instructors to artificially manufacture a stumbling block in
an in-class activity and demonstrate to students how to look up and
resolve a question. After developing some basic skills with
spreadsheets, students could be given a small problem to solve, which
requires some more creative thinking but it is not a huge conceptual
leap from what they've already seen. This allows students to build
foundational knowledge and a set of skills alongside problem-solving
strategies.

While much less common in K--12 education, we note that a number of
courses focused directly on spreadsheet tools exist in colleges and
universities \citep{bhcc, spc, agr-6932}. Typical topics often included
in such a course include:

\begin{itemize}
\tightlist
\item
  Basic terminology and key concepts
\item
  Formulas and functions
\item
  Formatting
\item
  Data management
\item
  Charting and graphing
\item
  Dealing with multiple sheets
\item
  Macros and programming
\item
  Data analysis tools
\item
  Data exchange
\end{itemize}

While this approach may be appropriate at the undergraduate level, we
concur with the advice of \citet{nasemk12:2026} that it is important to
integrate study of spreadsheet tools into a broader curriculum and not
just teach in a standalone elective course.

In any discussion of teaching technology in the present world, we need
to consider the impact of Generative AI \citep{ellis}. AI has the
potential to bring technology and programming skills closer to people
\citep{campbell_using_2024}, and allows individuals to pick up new
skills from whatever level they are at. Existing AI tools are often
effective when it comes to basic formula creation in spreadsheets
\citep{thorne_experimenting_2023}. For example, AI tutors may be a way
to support students coding using Apps Script.

Similar tools could also be helpful from a Professional Development (PD)
standpoint---instead of a school needing to engineer, say, a series of
PD courses depending on teacher background in Google Sheets, Generative
AI could be used to tailor information to a specific background and
complement more traditional PD modalities.

Of course, Generative AI also has the potential to be detrimental to
learning outcomes---for instance, by leading students to skip
foundational concepts and never develop a robust ``mental model'' of the
material. Further, these tools pose challenges for academic integrity,
as students may submit AI-generated work as their own; among the
possible responses, in-class activities that ask students to demonstrate
their learning in real time may be particularly useful. A thoughtful
conversation about how to address these merits and drawbacks would be
important when seeking to incorporate spreadsheets more deeply into the
curriculum.

When it comes to skills that students will develop, it is important to
somehow distinguish specific technical skills (such as the details for
how to apply conditional formatting, add dropdown menus, etc.) from more
general and transferable data science skills (e.g., being given a data
set and some questions to answer, then utilizing a spreadsheet to do
so). There is likely need and room for both types of elements in a K--12
curriculum. However, the exact balance between the two needs to be
determined.

Assessing learning success can be difficult and may take a variety of
different forms. Some approaches may include: written reports, trying to
see if students can use data and graphs to back up their claims;
blog-style articles describing findings from a spreadsheet project; or
in-class evaluations, where students work with their laptops out and
must complete some tasks in a spreadsheet.

In closing, we believe that spreadsheets have a part to play in K--12
data science education and can serve as a foundation for work in other
tools. The interactive and visually appealing elements of working with
spreadsheets have the potential to activate a creative side in students,
turning math or science from something they have to do to something fun
and interesting. This may allow for the engagement of an entirely
different group of students in STEM classes.

They play an important role as a component of a more general data
science toolbox. They have a role for data science practice
\citep{brom:woo:2018}. A recent survey of Master's graduates in
statistics reported that 69\% used Excel on a weekly or daily basis
\citep{asa-survey}. Anecdotal evidence \citep{agr-6932} suggests that
many graduate students lack sufficient spreadsheet skills.

We proposed five spreadsheet data skills that students should develop
and described how they might be integrated into a variety of courses.
Spreadsheet tools, when thoughtfully integrated into a K--12 curriculum
across science, math, and technology classes \citep{Nasir_2024}, can
help K--12 students develop data acumen. In a world increasingly
dependent on data, it is critical to ensure that students graduate from
high school equipped with the knowledge and resources that will allow
them to succeed throughout their lives---whether in further education
and/or professionally.

\section{Disclosure statement}\label{disclosure-statement}

The authors have no conflicts of interest to report. We thank the
editors and reviewers for their thoughtful and constructive feedback on
this manuscript, which has helped to improve the clarity and quality of
the paper. We gratefully acknowledge the contributions of teachers
Zachary Boboth (Sora Schools), Jed Dioguardi (Hackley School), Diana
Kaplan (Hackley School), Seth Karpinski (Hackley School), Daniel Lipin
(St.~Mark's School of Texas), Bill McLay (Hackley School), and Keshena
Richardson (Hackley School) for providing rich and insightful
discussions on effective strategies, challenges, and
considerations---both benefits and limitations---related to
incorporating spreadsheets into teaching and learning within the K--12
setting.

\section{Data Availability Statement}\label{data-availability-statement}

Resources and sample activities have been made available at the
following URL:
\href{https://docs.google.com/document/d/13L3h2NwN2wvn-v4-99yjkBgKSGR466IVyvw4-VHjGFE/edit?usp=sharing}{Google
Drive}

\newpage

\setcounter{section}{0}
\renewcommand{\thesection}{\Alph{section}}

\setcounter{table}{0}
\renewcommand{\thetable}{S\arabic{table}}

\setcounter{figure}{0}
\renewcommand{\thefigure}{S\arabic{figure}}

\section{Appendix (Code Examples)}\label{appendix-a}

As described in Section~\ref{sec-skills}, Google Apps Script is a
JavaScript-based programming language built into the Google Workspace
\citep{noauthor_launched_2009}. It is edited entirely in the browser,
with scripts executed on Google's servers---no software installation is
required. Apps Script is available with any Google account, including
education accounts.\footnote{Some Apps Script features may be restricted
  by a district's Google Workspace IT configuration.} Google also
provides an official quickstart guide for using Apps Script
\citep{gas_quickstart}.

There are two core ways in which Apps Script is used; these two use
cases speak to different goals in the classroom. \emph{Custom functions}
extend the functionality of Google Sheets. They are similar to the
formulas students are already familiar with, such as \texttt{AVERAGE()}
or \texttt{ROUND()}, but allow for more customized logic. In a Google
Sheet, students can create a custom function via ``Extensions''
\textgreater{} ``Apps Script''. There, a student could write a short
function called \texttt{genLetterGrade()} and then enter
\texttt{=genLetterGrade(0.85)} in a cell to convert a numerical score to
a letter grade.

\emph{Automation scripts} serve a different purpose. Rather than
returning a value, they carry out \emph{actions} across Google Workspace
services---for instance, a script might send emails through Gmail,
duplicate documents in Drive, or respond to submissions in Forms. These
scripts can be run manually from a menu or button, or launched
automatically through triggers tied to events (such as a form
submission) or schedules (such as a daily timer).

Example 1 demonstrates a custom function supporting algorithmic thinking
(Skill 4, Section~\ref{sec-skill4}). Examples 2--4 demonstrate
automation scripts that integrate multiple Google Workspace tools,
supporting Skill 5 (Section~\ref{sec-skill5}) (scripting, automation,
and tool integration).

\subsection{Example 1: Custom Letter Grade
Function}\label{example-1-custom-letter-grade-function}

This custom function converts a numerical grade to a letter grade using
conditional logic. Once saved in the Apps Script editor, it can be
called from any cell as \texttt{=genLetterGrade(A1)}.

\begin{Shaded}
\begin{Highlighting}[]
\ControlFlowTok{function} \FunctionTok{genLetterGrade}\NormalTok{(numGrade) \{}
  \ControlFlowTok{if}\NormalTok{ (numGrade }\SpecialCharTok{\textgreater{}} \FloatTok{0.9}\NormalTok{) \{}
    \FunctionTok{return}\NormalTok{(}\StringTok{"A"}\NormalTok{);}
\NormalTok{  \} }\ControlFlowTok{else} \ControlFlowTok{if}\NormalTok{ (numGrade }\SpecialCharTok{\textgreater{}} \FloatTok{0.8}\NormalTok{) \{}
    \FunctionTok{return}\NormalTok{(}\StringTok{"B"}\NormalTok{);}
\NormalTok{  \} }\ControlFlowTok{else} \ControlFlowTok{if}\NormalTok{ (numGrade }\SpecialCharTok{\textgreater{}} \FloatTok{0.7}\NormalTok{) \{}
    \FunctionTok{return}\NormalTok{(}\StringTok{"C"}\NormalTok{);}
\NormalTok{  \} }\ControlFlowTok{else}\NormalTok{ \{}
    \FunctionTok{return}\NormalTok{(}\StringTok{"Fail"}\NormalTok{);}
\NormalTok{  \}}
\NormalTok{\}}
\end{Highlighting}
\end{Shaded}

A more robust version of this function might incorporate input
validation---checking that the input is numeric and falls within a
plausible range. This kind of defensive programming is a useful habit to
develop \citep{McNamara02012018}, and provides practice with the data
consistency checking described in Skill 2 (Section~\ref{sec-skill2}).

\begin{Shaded}
\begin{Highlighting}[]
\ControlFlowTok{function} \FunctionTok{genLetterGrade}\NormalTok{(numGrade) \{}
  \ControlFlowTok{if}\NormalTok{ (typeof numGrade }\SpecialCharTok{!=}\ErrorTok{=} \StringTok{"number"} \SpecialCharTok{||} \FunctionTok{isNaN}\NormalTok{(numGrade)) \{}
    \FunctionTok{return}\NormalTok{(}\StringTok{"Error: input must be a number"}\NormalTok{);}
\NormalTok{  \}}
  \ControlFlowTok{if}\NormalTok{ (numGrade }\SpecialCharTok{\textless{}} \DecValTok{0} \SpecialCharTok{||}\NormalTok{ numGrade }\SpecialCharTok{\textgreater{}} \DecValTok{1}\NormalTok{) \{}
    \FunctionTok{return}\NormalTok{(}\StringTok{"Error: input must be between 0 and 1"}\NormalTok{);}
\NormalTok{  \}}
  \ControlFlowTok{if}\NormalTok{ (numGrade }\SpecialCharTok{\textgreater{}} \FloatTok{0.9}\NormalTok{) \{}
    \FunctionTok{return}\NormalTok{(}\StringTok{"A"}\NormalTok{);}
\NormalTok{  \} }\ControlFlowTok{else} \ControlFlowTok{if}\NormalTok{ (numGrade }\SpecialCharTok{\textgreater{}} \FloatTok{0.8}\NormalTok{) \{}
    \FunctionTok{return}\NormalTok{(}\StringTok{"B"}\NormalTok{);}
\NormalTok{  \} }\ControlFlowTok{else} \ControlFlowTok{if}\NormalTok{ (numGrade }\SpecialCharTok{\textgreater{}} \FloatTok{0.7}\NormalTok{) \{}
    \FunctionTok{return}\NormalTok{(}\StringTok{"C"}\NormalTok{);}
\NormalTok{  \} }\ControlFlowTok{else}\NormalTok{ \{}
    \FunctionTok{return}\NormalTok{(}\StringTok{"Fail"}\NormalTok{);}
\NormalTok{  \}}
\NormalTok{\}}
\end{Highlighting}
\end{Shaded}

\subsection{Example 2: Automating File
Duplication}\label{example-2-automating-file-duplication}

This script automates the duplication and renaming of a Google Docs
template. Given a template document URL, it creates one copy for each
weekday, naming each copy after the corresponding day. Here the function
does not return a value but instead performs its work as a side effect
when run.

\begin{Shaded}
\begin{Highlighting}[]
\ControlFlowTok{function} \FunctionTok{duplicateNotesFile}\NormalTok{() \{}
  \SpecialCharTok{/}\ErrorTok{/}\NormalTok{ User}\SpecialCharTok{:}\NormalTok{ add your document URL here}
\NormalTok{  const templateDocUrl }\OtherTok{=} \StringTok{\textquotesingle{}YOUR URL HERE\textquotesingle{}}\NormalTok{;}

\NormalTok{  const docId }\OtherTok{=} \FunctionTok{DocumentApp.openByUrl}\NormalTok{(templateDocUrl)}\FunctionTok{.getId}\NormalTok{();}
\NormalTok{  const docFile }\OtherTok{=} \FunctionTok{DriveApp.getFileById}\NormalTok{(docId);}

  \SpecialCharTok{/}\ErrorTok{/}\NormalTok{ What folder should the template copies go }\ControlFlowTok{in}\NormalTok{?}
  \SpecialCharTok{/}\ErrorTok{/}\NormalTok{ This code places them }\ControlFlowTok{in}\NormalTok{ the same folder as the template.}
\NormalTok{  const destinationFolder }\OtherTok{=} \FunctionTok{docFile.getParents}\NormalTok{()}\FunctionTok{.next}\NormalTok{();}

\NormalTok{  const weekdays }\OtherTok{=}\NormalTok{ [}\StringTok{\textquotesingle{}Mon\textquotesingle{}}\NormalTok{, }\StringTok{\textquotesingle{}Tues\textquotesingle{}}\NormalTok{, }\StringTok{\textquotesingle{}Wed\textquotesingle{}}\NormalTok{, }\StringTok{\textquotesingle{}Thurs\textquotesingle{}}\NormalTok{, }\StringTok{\textquotesingle{}Fri\textquotesingle{}}\NormalTok{];}

  \ControlFlowTok{for}\NormalTok{ (let }\AttributeTok{i =} \DecValTok{0}\NormalTok{; i }\SpecialCharTok{\textless{}}\NormalTok{ weekdays.length; i}\SpecialCharTok{++}\NormalTok{) \{}
\NormalTok{    let weekday }\OtherTok{=}\NormalTok{ weekdays[i];}
\NormalTok{    let fileNm }\OtherTok{=}\NormalTok{ weekday }\SpecialCharTok{+} \StringTok{\textquotesingle{} Notes\textquotesingle{}}\NormalTok{;}
    \FunctionTok{docFile.makeCopy}\NormalTok{(fileNm, destinationFolder);}
    \FunctionTok{console.log}\NormalTok{(}\StringTok{\textquotesingle{}Successfully created file \textquotesingle{}} \SpecialCharTok{+}\NormalTok{ fileNm);}
\NormalTok{  \}}
\NormalTok{\}}
\end{Highlighting}
\end{Shaded}

\subsection{Example 3: Form Response
Notifications}\label{example-3-form-response-notifications}

This script, bound to a Google Form, sends email notifications when the
number of responses reaches specified thresholds (e.g., 10, 15, or 20
submissions).

\begin{Shaded}
\begin{Highlighting}[]
\SpecialCharTok{/}\ErrorTok{/}\NormalTok{ Run this }\ControlFlowTok{function}\NormalTok{ first, to create a }\StringTok{"trigger"}\NormalTok{, thereby having the main}
\SpecialCharTok{/}\ErrorTok{/}\NormalTok{ code run every time a form response is submitted.}
\SpecialCharTok{/}\ErrorTok{/}\NormalTok{ This can also be done through the UI.}
\ControlFlowTok{function} \FunctionTok{makeTrigger}\NormalTok{() \{}
  \FunctionTok{ScriptApp.newTrigger}\NormalTok{(}\StringTok{\textquotesingle{}checkResponseFreq\textquotesingle{}}\NormalTok{)}
    \FunctionTok{.forForm}\NormalTok{(}\FunctionTok{FormApp.getActiveForm}\NormalTok{())}
    \FunctionTok{.onFormSubmit}\NormalTok{()}
    \FunctionTok{.create}\NormalTok{();}
\NormalTok{\}}

\ControlFlowTok{function} \FunctionTok{checkResponseFreq}\NormalTok{() \{}
\NormalTok{  const notificationQuantities }\OtherTok{=}\NormalTok{ [}\DecValTok{10}\NormalTok{, }\DecValTok{15}\NormalTok{, }\DecValTok{20}\NormalTok{];}
\NormalTok{  const notificationEmails }\OtherTok{=}\NormalTok{ [}\StringTok{\textquotesingle{}test123@gmail.com\textquotesingle{}}\NormalTok{, }\StringTok{\textquotesingle{}testabc@gmail.com\textquotesingle{}}\NormalTok{];}

\NormalTok{  const formResponses }\OtherTok{=} \FunctionTok{FormApp.getActiveForm}\NormalTok{()}\FunctionTok{.getResponses}\NormalTok{();}

  \ControlFlowTok{if}\NormalTok{ (}\FunctionTok{notificationQuantities.includes}\NormalTok{(formResponses.length)) \{}
    \FunctionTok{GmailApp.sendEmail}\NormalTok{(}
\NormalTok{      notificationEmails,}
      \StringTok{\textquotesingle{}Form submission update\textquotesingle{}}\NormalTok{,}
      \StringTok{\textasciigrave{}}\AttributeTok{This is a notification informing you that $\{formResponses.length\}}
\AttributeTok{      form responses have been submitted thus far.}\StringTok{\textasciigrave{}}
\NormalTok{    );}

    \FunctionTok{console.log}\NormalTok{(}\StringTok{"Sent notification email"}\NormalTok{)}
\NormalTok{  \}}
\NormalTok{\}}
\end{Highlighting}
\end{Shaded}

\subsection{Example 4: Deadline Reminder
Emails}\label{example-4-deadline-reminder-emails}

This script, bound to a Google Sheets spreadsheet, sends daily email
reminders for upcoming deadlines. It checks each row of the tracking
spreadsheet and sends a reminder if the deadline is within a specified
number of days and the task has not yet been marked complete.

\begin{Shaded}
\begin{Highlighting}[]
\SpecialCharTok{/}\ErrorTok{/}\NormalTok{ Run this }\ControlFlowTok{function}\NormalTok{ first, to create a }\StringTok{"trigger"}\NormalTok{, thereby having the main}
\SpecialCharTok{/}\ErrorTok{/}\NormalTok{ code run every day at around }\DecValTok{8}\NormalTok{am. This can also be done through the UI.}
\ControlFlowTok{function} \FunctionTok{makeTrigger}\NormalTok{() \{}
  \FunctionTok{ScriptApp.newTrigger}\NormalTok{(}\StringTok{\textquotesingle{}sendReminderEmail\textquotesingle{}}\NormalTok{)}
    \FunctionTok{.timeBased}\NormalTok{()}
    \FunctionTok{.atHour}\NormalTok{(}\DecValTok{8}\NormalTok{)}
    \FunctionTok{.nearMinute}\NormalTok{(}\DecValTok{0}\NormalTok{)}
    \FunctionTok{.everyDays}\NormalTok{(}\DecValTok{1}\NormalTok{)}
    \FunctionTok{.create}\NormalTok{();}
\NormalTok{\}}

\ControlFlowTok{function} \FunctionTok{sendReminderEmail}\NormalTok{() \{}
\NormalTok{  const sheetCols }\OtherTok{=}\NormalTok{ \{}
    \StringTok{\textquotesingle{}company\textquotesingle{}}\SpecialCharTok{:} \DecValTok{0}\NormalTok{,}
    \StringTok{\textquotesingle{}dueDate\textquotesingle{}}\SpecialCharTok{:} \DecValTok{1}\NormalTok{,}
    \StringTok{\textquotesingle{}complete\textquotesingle{}}\SpecialCharTok{:} \DecValTok{2}\NormalTok{,}
\NormalTok{  \}}

  \SpecialCharTok{/}\ErrorTok{/}\NormalTok{ By default, send emails to the person who created the trigger}
\NormalTok{  const reminderRecipient }\OtherTok{=} \FunctionTok{Session.getEffectiveUser}\NormalTok{()}\FunctionTok{.getEmail}\NormalTok{();}

  \SpecialCharTok{/}\ErrorTok{/}\NormalTok{ By default, have the reminder emails sent seven days }\ControlFlowTok{in}\NormalTok{ advance.}
\NormalTok{  const reminderNDays }\OtherTok{=} \DecValTok{7}\NormalTok{;}

\NormalTok{  const sSht }\OtherTok{=} \FunctionTok{SpreadsheetApp.getActiveSpreadsheet}\NormalTok{();}

  \SpecialCharTok{/}\ErrorTok{/}\NormalTok{ Assumes data will be on the first tab of the spreadsheet}
\NormalTok{  const sheet }\OtherTok{=} \FunctionTok{sSht.getSheets}\NormalTok{()[}\DecValTok{0}\NormalTok{];}

\NormalTok{  const vals }\OtherTok{=} \FunctionTok{sheet.getDataRange}\NormalTok{()}\FunctionTok{.getValues}\NormalTok{();}

  \SpecialCharTok{/}\ErrorTok{/}\NormalTok{ Remove header row}
  \FunctionTok{vals.shift}\NormalTok{();}

  \ControlFlowTok{for}\NormalTok{ (let row of vals) \{}
\NormalTok{    let company }\OtherTok{=}\NormalTok{ row[sheetCols.company];}
\NormalTok{    let dueDate }\OtherTok{=}\NormalTok{ row[sheetCols.dueDate];}
\NormalTok{    let complete }\OtherTok{=}\NormalTok{ row[sheetCols.complete];}
    \FunctionTok{console.log}\NormalTok{(}\StringTok{\textasciigrave{}}\AttributeTok{Application for $\{company\} is due on $\{dueDate\};}
\AttributeTok{                completion status $\{complete\}}\StringTok{\textasciigrave{}}\NormalTok{);}

    \ControlFlowTok{if}\NormalTok{ (}\FunctionTok{isNDaysAway}\NormalTok{(dueDate, reminderNDays) }\SpecialCharTok{\&\&} \SpecialCharTok{!}\NormalTok{complete) \{}
      \FunctionTok{GmailApp.sendEmail}\NormalTok{(}
\NormalTok{        reminderRecipient,}
        \StringTok{\textquotesingle{}Upcoming job application deadline\textquotesingle{}}\NormalTok{,}
        \StringTok{\textasciigrave{}}\AttributeTok{This is a reminder that your application to $\{company\} is}
\AttributeTok{        $\{reminderNDays\} days away, and you have not yet submitted it!}\StringTok{\textasciigrave{}}\NormalTok{);}
\NormalTok{    \}}
\NormalTok{  \}}
\NormalTok{\}}

\SpecialCharTok{/}\ErrorTok{/}\NormalTok{ A }\ControlFlowTok{function}\NormalTok{ to check }\ControlFlowTok{if}\NormalTok{ a given date is}
\SpecialCharTok{/}\ErrorTok{/}\NormalTok{ exactly }\StringTok{\textasciigrave{}}\AttributeTok{nDays}\StringTok{\textasciigrave{}}\NormalTok{ away from the current date.}
\ControlFlowTok{function} \FunctionTok{isNDaysAway}\NormalTok{(date, nDays) \{}
\NormalTok{  const now }\OtherTok{=}\NormalTok{ new }\FunctionTok{Date}\NormalTok{();}
\NormalTok{  const futureDate }\OtherTok{=}\NormalTok{ new }\FunctionTok{Date}\NormalTok{(}\FunctionTok{now.getTime}\NormalTok{() }\SpecialCharTok{+}\NormalTok{ nDays }\SpecialCharTok{*} \DecValTok{24} \SpecialCharTok{*} \DecValTok{60} \SpecialCharTok{*} \DecValTok{60} \SpecialCharTok{*} \DecValTok{1000}\NormalTok{);}

\NormalTok{  return }\FunctionTok{date.toDateString}\NormalTok{() }\SpecialCharTok{==}\ErrorTok{=} \FunctionTok{futureDate.toDateString}\NormalTok{();}
\NormalTok{\}}
\end{Highlighting}
\end{Shaded}

\bibliography{bibliography.bib}

\end{document}